\documentclass[letterpaper, journal, twoside]{IEEEtran}

\IEEEoverridecommandlockouts

\usepackage{dsfont}
\usepackage[T1]{fontenc}
\usepackage{graphicx, stfloats} 
\usepackage{svg}
\usepackage{geometry}
\usepackage{standalone}
\usepackage{hyperref}
\usepackage{pgfplots}
\usepackage{booktabs}
\usepackage{subcaption}
\usepackage{amssymb, amsmath}
\usepackage{algorithm}
\usepackage{algpseudocode}
\usepackage{makecell}
\usepackage{bbm}
\usepackage{diagbox, multirow} 
\usepackage{color,soul} 
\usepackage{siunitx}

\definecolor{vandeusen}{RGB}{73,92,111}
\definecolor{cordovan}{RGB}{152,68,71}
\definecolor{alizarin}{rgb}{0.82, 0.1, 0.26}
\definecolor{azure}{rgb}{0.0, 0.5, 1.0}

\title{Hydrogen in Aviation: Evaluating the Feasibility and Benefits of a Green Fuel Alternative}

\author{ Armaan Sharma$^{1}$, Mansur M. Arief$^{2\star}$%

\thanks{$^1$Armaan Sharma is with the Hillsborough High School, Hillsborough, NJ, USA}%
\thanks{$^2$Mansur M. Arief is with the Department of Aeronautics and Astronautics, Stanford University, Stanford, CA, USA (\url{mansur.arief@stanford.edu})}%
\thanks{$^{\star}$Corresponding Author}%
}

\begin{document}

\markboth{Preprint Version. December, 2024}
{Sharma \& Arief (2024): Hydrogen in Aviation: Evaluating the Feasibility and Benefits of a Green Fuel Alternative}

\maketitle

\begin{abstract}
Growing concerns regarding environmental health have highlighted the aviation industry's impact and potential mitigation strategies. Previous research has indicated hydrogen's significant potential for reducing the industry's environmental impact, yet implementation challenges remain. Through analysis of light aircraft and military applications, we demonstrate that hydrogen-based systems can achieve performance metrics approaching those of traditional fuels while reducing emissions by up to 74.7\%. Our findings show that hydrogen's superior energy-to-mass ratio (120 MJ/kg versus 43 MJ/kg for jet fuel) makes it particularly advantageous for aviation applications compared to battery-electric alternatives. Primary implementation challenges involve cryogenic storage systems (-253°C), tank placement optimization, and fueling infrastructure development. The observed efficiency penalties of only 2.23\% in military applications suggest hydrogen's viability as a sustainable aviation fuel alternative.
\end{abstract}


\section{Introduction}

\IEEEPARstart{H}{umans'} use of detrimental fuel types is causing environmental conditions to change rapidly. Society has grown to depend on fossil fuels as a crucial source of energy used worldwide \cite{surer2018state}. In turn, fossil fuels have become more affordable and easily accessible. However, these fuels damage the environment and deliver a bleak conception of the planet in the long term. The combustion of fossil fuels releases greenhouse gases, primarily carbon dioxide, which contribute to global climate change by trapping heat in the Earth's atmosphere. Consequently, this leads to rising global temperatures, extreme weather events, and degradation of air quality, affecting public health through increased rates of respiratory diseases, cardiovascular conditions, and premature mortality \cite{anwar2015causes}. Studies have shown that toxic air quality causes a myriad of health issues which range from relatively minor conditions like asthma to significant ones like lung cancer or death \cite{behinaein2023growing}. To address these problems, sustainable sources of energy must be analyzed.

\begin{figure}
    \centering
    \includegraphics[width=\linewidth]{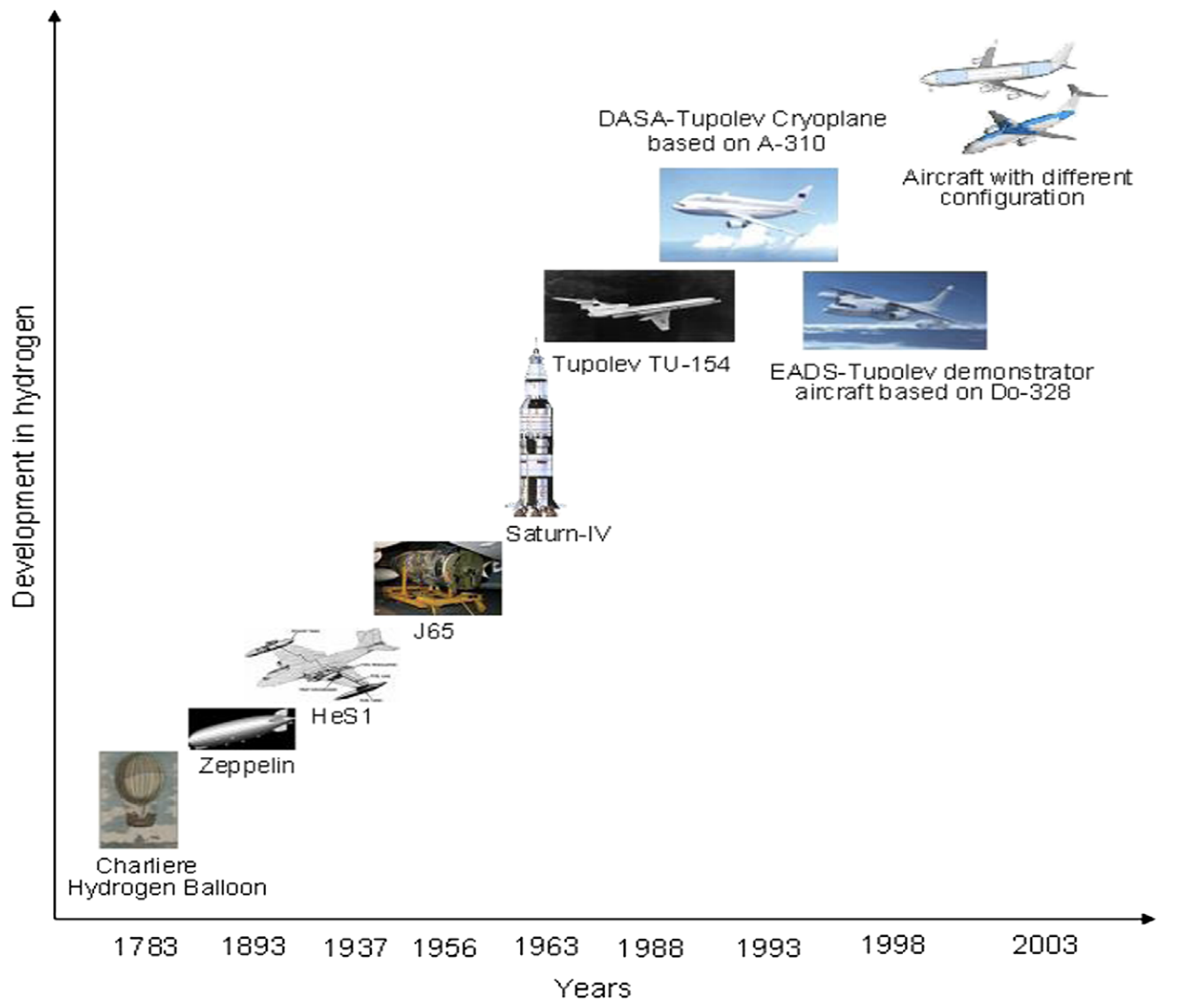}
    \caption{Historical development timescale for hydrogen \cite{surer2018state}}
    \label{fig:historical_dev}
\end{figure}

One major contributor to the environmental problem is the aviation industry, which, from manufacturing, operations, and maintenance, is responsible for 12\% of all harmful gas emissions caused by the transportation sector \cite{surer2018state}. While manufacturing processes present their own environmental challenges, this study focuses specifically on operational emissions and their potential reduction through alternative fuel sources. These emissions pose a particular challenge as air traffic continues to grow globally. 

Among various alternative fuel options, hydrogen emerges as a particularly promising solution for aviation due to its unique combination of high energy density by mass, zero direct carbon emissions, and potential for sustainable production through renewable energy sources. Many solutions to this problem have been explored, but hydrogen stands out as a promising alternative for powering airplanes. Hydrogen is an optimal fuel source for multiple reasons. First, when used as an energy source, its only emission is water vapor \cite{surer2018state}, therefore not damaging any part of the environment. Second, it is abundant in nature. The hydrogen gas can be formed by breaking up water molecules and connecting hydrogen atoms. Additionally, hydrogen can be found naturally. The gas forms underground when water reacts with minerals. Because this process is ongoing, natural hydrogen is renewable \cite{usgsgeologic2024}. Unlike manufactured hydrogen, hydrogen made naturally does not require processes that use fossil fuels. Finally, hydrogen is more efficient than electricity in a battery for storage. Batteries are heavy and require specific storage areas due to their restrictive shapes, while hydrogen tanks can be placed in many locations, such as above an aircraft's ceiling \cite{yusaf2024sustainable}.

The viability of hydrogen as an alternative fuel source has been demonstrated across multiple transportation sectors, most notably in automotive applications where fuel cell vehicles have achieved commercial deployment. The aviation sector has also shown promising early results, with several successful demonstrations dating back to the mid-20th century. These include the hydrogen-powered flights of the U.S. Air Force's Martin B57 Canberra and the Soviet Union's Tupolev Tu-154, which provided valuable data on hydrogen fuel systems in aerospace applications. As illustrated in \autoref{fig:historical_dev}, technological development has continued steadily, with recent advances in fuel cell efficiency, storage systems, and safety protocols bringing hydrogen-powered aviation closer to practical implementation.

This paper presents a systematic evaluation of hydrogen's potential as an aviation fuel through analysis of its technical characteristics, implementation challenges, and performance in two key applications: light aircraft and military aviation. Our investigation particularly focuses on the efficiency, emissions reduction potential, and practical feasibility of hydrogen fuel systems compared to conventional aviation fuels. The remainder of this paper is organized as follows. In Section~\ref{sec:related_work}, we present a literature review of alternative aviation fuels and existing hydrogen applications, providing essential context for our analysis. Section~\ref{sec:framework} describes our analytical framework and methodology. In Section~\ref{sec:case_studies}, we examine detailed case studies of hydrogen implementation. Section~\ref{sec:discussion} analyzes the implications of our findings, and Section~\ref{sec:conclusion} presents conclusions and future research directions.

\section{Literature Review}\label{sec:related_work}
This review examines two critical areas that inform the potential of hydrogen in aviation: the current state and limitations of alternative aviation fuels, and existing applications of hydrogen fuel systems. This structure provides both theoretical context and practical evidence for evaluating hydrogen's viability in aerospace applications.

\subsection{Alternative Fuels}

\begin{figure}
    \centering
    \includegraphics[width=\linewidth]{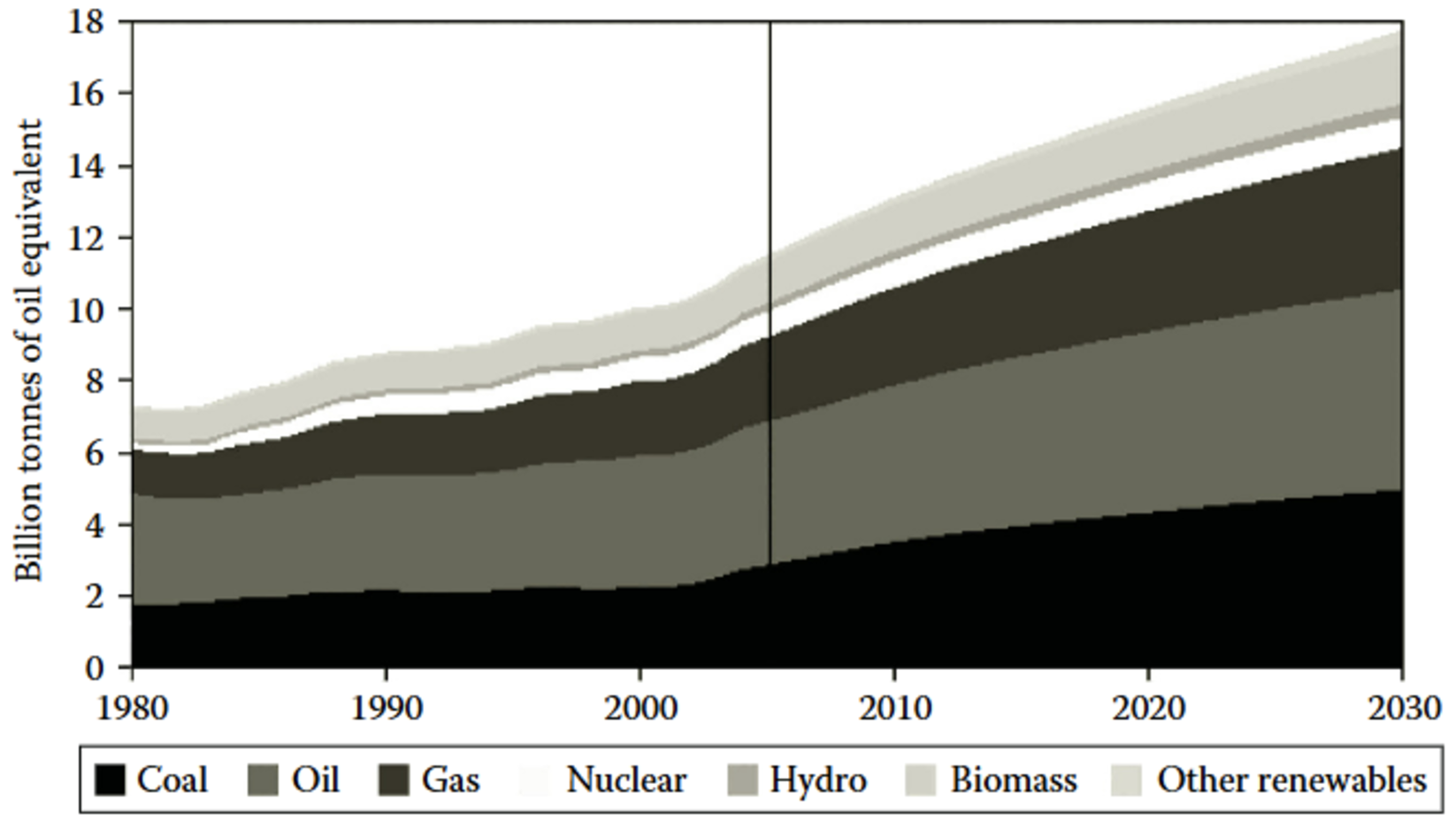}
    \caption{Recent and projected rises in fuel usage \cite{ramadhas2011alternative}}
    \label{fig:projected_fuel_usage}
\end{figure}

The aviation industry's dependence on fossil fuels presents both environmental and sustainability challenges that necessitate the exploration of alternative fuel sources. Current fuel consumption patterns show concerning trends, with fossil fuel usage rising sharply in recent years. Their importance has grown to such an extent that they can determine the economic strength of a country, with economies around the world now deeply dependent on fossil fuels \cite{ramadhas2011alternative}. As shown in \autoref{fig:projected_fuel_usage}, the usage of coal, oil, and gas has risen over the past few decades and is projected to continue increasing.

This trajectory poses significant environmental challenges, particularly in terms of greenhouse gas emissions. Projections indicate that by 2035, aviation-related emissions could double compared to 2007 levels \cite{ramadhas2011alternative}. This increase would significantly impact global climate change mitigation efforts and air quality, particularly around aviation hubs.

For alternative fuels to effectively replace conventional aviation fuels, they must meet several critical criteria: adequate energy density for aviation applications, practical storage and transport capabilities, economic viability, and compatibility with existing or modified aircraft systems \cite{ramadhas2011alternative}. Current alternatives under development include biofuels and compressed natural gas, which have shown promise in terms of production scalability and cost-effectiveness.

Among these alternatives, hydrogen emerges as particularly promising due to its unique characteristics. When used in fuel cells or direct combustion, hydrogen produces only water vapor as a byproduct, offering a potentially carbon-neutral energy solution. However, the environmental impact of hydrogen varies significantly based on its production method. Current production relies heavily on fossil fuels, with 48\% derived from natural gas, 30\% from oils, and 18\% from coal~\cite{nikolaidis2017comparative}.

The development of greener hydrogen production methods represents a crucial advancement toward truly sustainable aviation. These include biomass-derived hydrogen, algae-based production through photosynthesis, and most promisingly, electrolysis powered by renewable energy. While electrolysis currently presents higher production costs compared to fossil fuel-based methods, it offers the most environmentally sustainable pathway for hydrogen production~\cite{nikolaidis2017comparative}.

\subsection{Existing Applications}
Practical implementations of hydrogen fuel systems in aviation have demonstrated both the potential and challenges of this technology. Light aircraft applications have shown particularly promising results. In a study completed by British-American hydrogen aircraft development company Zeroavia, a 14-passenger Cessna 208 Caravan turboprop aircraft was retrofitted with a liquid hydrogen-fueled fuel-cell system. The airplane successfully completed a simulated 350-kilometer (~218-mile) flight that lasted approximately 1.5 hours under normal flight conditions and power settings \cite{kasim2022performance}. Additionally, successful demonstrations in urban air mobility applications, such as the hydrogen fuel cell-powered air taxi in Stuttgart, Germany, which achieved speeds up to 200 kilometers per hour, suggest the technology's viability for short-range commercial applications \cite{surer2018state}.

Military applications have provided valuable data on hydrogen fuel systems under more demanding conditions. Early demonstrations, including the United States Air Force's successful operation of a hydrogen-powered Martin B57 Canberra bomber in 1956 \cite{surer2018state} and the Soviet Union's tests with a modified Tupolev Tu-154 airliner in 1988 \cite{surer2018state}, established basic feasibility. More recent developments, such as Boeing's Phantom Eye, a high-altitude and long-range reconnaissance aircraft powered entirely by liquid hydrogen, have demonstrated the potential advantages of hydrogen in specialized military applications, particularly for missions requiring extended endurance \cite{mills2012design}.

These implementations across both civilian and military sectors have provided crucial insights into the practical challenges and opportunities of hydrogen aviation. The progression of hydrogen technology in aviation, as illustrated in \autoref{fig:historical_dev}, shows steady advancement in both technical capabilities and operational understanding. However, key challenges remain in areas such as fuel storage, distribution infrastructure, and system integration, particularly for larger aircraft applications.

\section{Framework}\label{sec:framework}

This study employs a comparative analysis framework to evaluate hydrogen's viability as an aviation fuel source. Our methodology focuses on three key performance metrics: energy density characteristics, storage requirements, and system efficiency. This approach enables systematic comparison between hydrogen, conventional jet fuel, and battery-electric alternatives.

The first critical metric is power-to-weight ratio, which directly impacts aircraft performance and range. Military aircraft require fuels with exceptionally high energy densities to meet their high-performance demands. Similarly, commercial airliners and cargo planes depend on high-density fuels since they transport heavy payloads over long distances and cannot afford to carry unnecessary weight. 

\begin{figure}
    \centering
    \includegraphics[width=\linewidth]{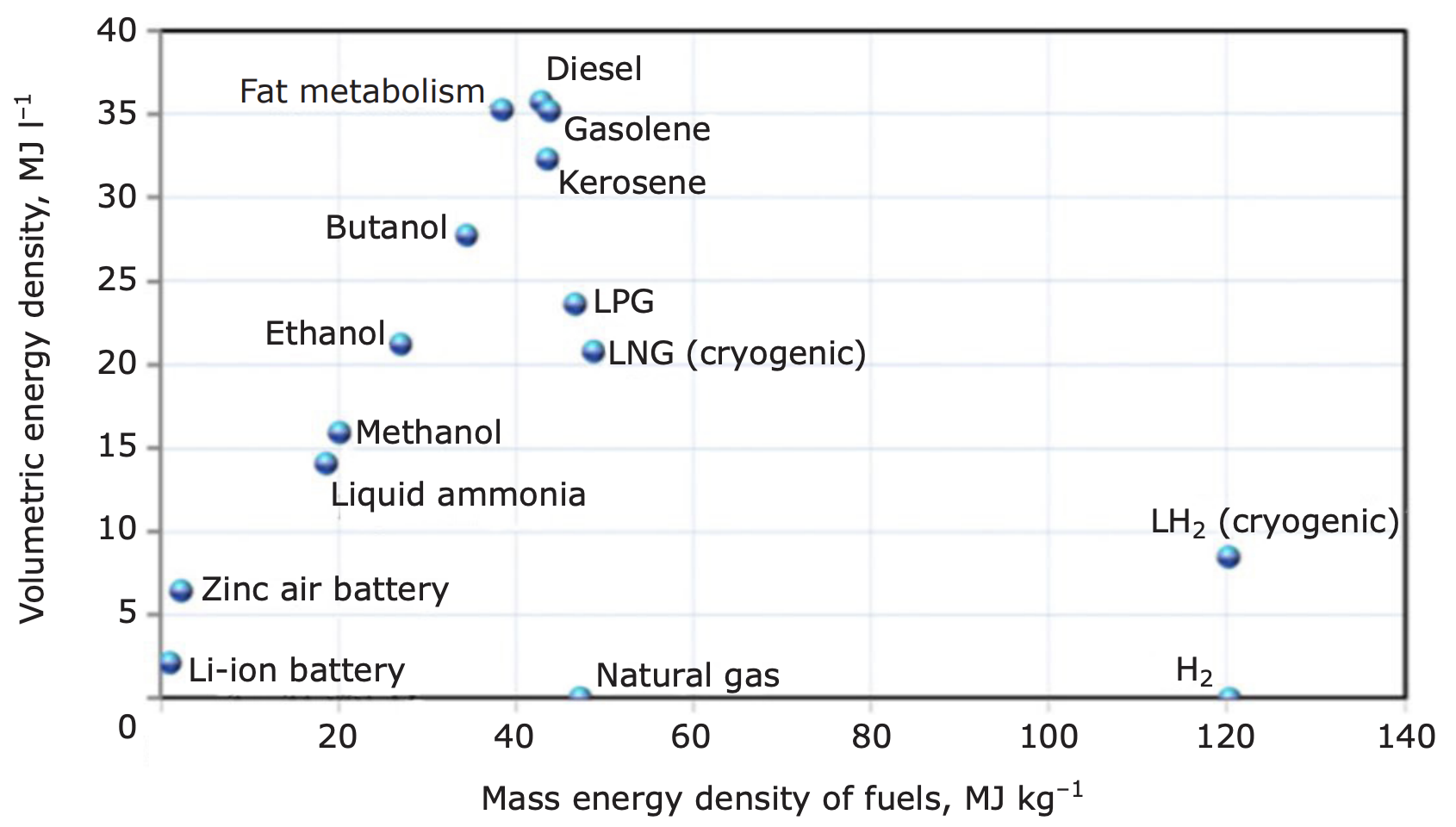}
    \caption{Comparison of various energy sources for aviation~\cite{bauen2020sustainable}}
    \label{fig:energy_source_companrison}
\end{figure}

Quantitative analysis demonstrates that while hydrogen has a volumetric energy density approximately one-quarter that of traditional kerosene jet fuel (8.9 MJ/L versus 35 MJ/L), its gravimetric energy density is significantly higher (120 MJ/kg versus 43 MJ/kg) \cite{adler2023hydrogen, bauen2020sustainable}, as illustrated in \autoref{fig:energy_source_companrison}. This characteristic provides a crucial advantage in aviation applications where weight minimization is paramount. For comparison, lithium-ion batteries achieve only 0.9 MJ/kg, highlighting hydrogen's substantial advantage over battery-electric systems for aviation applications.

The second key metric examines storage system requirements and their impact on aircraft design and operation. Airplanes are designed with aerodynamics as a key concern, using slim fuselages and wings to optimize their airflow. Traditional jet fuel storage benefits from the fuel's liquid state at ambient temperatures and its ability to be stored in wing tanks, effectively utilizing otherwise empty space while maintaining aerodynamic efficiency.

Hydrogen storage presents distinct challenges due to its physical properties. The two primary storage approaches each present unique considerations for aircraft design. Compressed gaseous hydrogen requires high-pressure tanks operating at 350-700 bar, necessitating additional weight from tank materials and safety systems. Alternatively, liquid hydrogen storage demands cryogenic conditions at -253°C, requiring sophisticated insulation systems and accounting for boil-off losses during operation.

These storage requirements significantly influence aircraft design considerations. The placement of storage tanks must be carefully optimized to maintain proper center of gravity while minimizing aerodynamic impact. The lower volumetric density of hydrogen necessitates approximately four times more volume compared to conventional fuel storage for equivalent energy content. Furthermore, aircraft structures may require modifications to accommodate cryogenic systems and meet additional safety requirements associated with hydrogen fuel systems.

The storage challenges become particularly acute in small unmanned aircraft, where space and weight constraints are severe. However, for commercial and military aircraft, hydrogen storage systems offer advantages over battery installations. The cylindrical geometry of hydrogen tanks aligns well with fuselage design, whereas batteries' rectangular form factors present greater integration challenges.

The third metric considers system-level efficiency through multiple interconnected factors. Hydrogen fuel cells typically operate at 45-60\% efficiency, with additional considerations for power system integration losses and the storage system's impact on overall aircraft performance. The combination of these factors determines the practical efficiency of hydrogen-powered aircraft systems.

This comprehensive framework provides a systematic basis for evaluating hydrogen's potential in specific aviation applications, as demonstrated in the following case studies. The analysis considers both technical performance metrics and practical implementation challenges, enabling a thorough assessment of hydrogen's viability as an aviation fuel alternative. By examining these three key metrics -- energy density, storage requirements, and system efficiency -- we can evaluate the practical feasibility of hydrogen fuel systems across different aviation applications.

\section{Case Studies}\label{sec:case_studies}

Through analysis of two distinct applications, we evaluate the practical implementation and performance characteristics of hydrogen fuel systems in aviation contexts. These cases examine both light aircraft and military applications, providing insights into hydrogen's versatility across different operational requirements.

\subsection{Light Aircraft}

The implementation of hydrogen fuel systems in light aircraft provides valuable data on performance comparisons with conventional aviation fuels. A notable example is a case study that compared the performance of a hydrogen fuel cell system-powered Novotech Seagull, an amphibious two-seat light aircraft, against a conventional aviation fuel-powered Seagull in simulated flights. Both simulations replicated typical flights of aircraft in this class–a takeoff and climb to approximately 2,500 feet, a short cruise, a further climb to 6,000 feet, a longer cruise, and finally a descent and landing. 

The fuel cell-powered Seagull was tested with both a fuel cell system and a hybrid-electric system in which batteries could be charged during the descent portion. The takeoff phase was crucial; it determined the fuel cell system's acceleration compared to the conventional system, thus demonstrating each system's maximum power and capacity. The propulsion systems' fuel efficiencies and power figures were compared at multiple points throughout the flight to show both the fuel cell's and aviation fuel's benefits and drawbacks at various altitudes. The study provided quantitative data on emissions reduction potential, a key metric for evaluating environmental impact.

The data revealed striking differences: nitrogen oxide and carbon monoxide gas emissions from the aviation fuel-powered plane substantially exceeded those from the hydrogen-powered plane with the battery-charging technology. In sample flights, the gas-powered plane emitted approximately 87 grams of nitrogen oxide and 15 grams of carbon monoxide, while the fuel cell-equipped plane emitted only 22 grams of nitrogen oxide and 10 grams of carbon monoxide when fueled with hydrogen made from natural gas. These results demonstrate significant emissions reductions even when using hydrogen produced through conventional methods.

The analysis of hybrid system performance revealed important trade-offs in system design. The battery system increases hydrogen consumption due to its weight, partially diminishing the emission benefits. Nevertheless, this arrangement provides vital backup power in case of hydrogen system malfunction. If batteries were exclusively used, their considerable weight would nullify any desired efficiency improvements. In a hybrid propulsion system, the batteries are not as large as they would be in a full-electric scenario, which results in a reduced impact on weight. For this reason, a hybrid system in which hydrogen serves as the main energy source and a battery provides a backup source remains the most advantageous propulsion solution for small aircraft \cite{donateo2024energy}.

\subsection{Military Aircraft}

Military aviation applications provide insights into hydrogen's performance under more demanding operational conditions. Similar to its applications in light aircraft, hydrogen can effectively power larger military aircraft. This was demonstrated in a study in which the General Electric J85-GE-5H turbojet engine was experimentally powered with both conventional JP-8 fuel, a common fuel for turbojet engines, and hydrogen. This engine is used to power the Northrop T-38 Talon supersonic military training aircraft. Among the statistics gathered in the study, exergetic efficiency—a measure of wasted energy and the overall performance of a system—provided a comparison between the effectiveness of traditional fuels and hydrogen. 

With traditional JP-8 fuel, the exergetic efficiency was 30.85\% while the turbojet engine was not using its afterburner and 16.98\% with the afterburner. Hydrogen yielded minor efficiency losses, with the exergetic efficiency being 28.62\% without the afterburner and 15.33\% with the afterburner. These modest deficits were offset by the observed reduction of greenhouse gas emissions, which fulfilled the goal of benefiting the environment \cite{yuksel2020comparative}. The relatively small efficiency differences demonstrate hydrogen's viability for high-performance military applications.

Most fighter aircraft designs prioritize minimal weight and drag-inducing elements to sustain flight at high altitudes and speeds. A heavy battery would compromise this requirement, leading to an efficiency decline during the most demanding stages of an air mission. As previously discussed, hydrogen's ratio of energy to mass is much greater than that of most types of batteries. The weight reduction hydrogen provides enables an efficiency gain. Additionally, batteries' inefficient configurations would necessitate a departure from fighter airplanes' small and narrow drag-reducing designs. Hydrogen's form-fitting storage tanks and lightweight properties establish it as the ideal green power source for military aviation.

\section{Discussion}\label{sec:discussion}

Our analysis of hydrogen fuel systems in aviation applications reveals both promising capabilities and specific implementation challenges. The comparative evaluation of hydrogen against conventional jet fuel and battery-electric systems can be summarized across three critical performance criteria, as shown in \autoref{tbl:summary}.

\begin{table}[]
\caption{Comparative analysis of aviation fuel alternatives}
\setlength{\tabcolsep}{11pt}
\begin{tabular}{@{}llll@{}}
\toprule
\multicolumn{1}{c}{\multirow{2}{*}{\textbf{Performance Criteria}}} & \multicolumn{3}{c}{\textbf{Fuel System Type}}                     \\ \cmidrule(l){2-4} 
\multicolumn{1}{c}{}                                   & \textbf{Hydrogen} & \textbf{Jet Fuel} & \textbf{Batteries} \\ \midrule
System Efficiency                                      & Good              & Best              & Worst              \\
Environmental Impact                                   & Best              & Worst             & Good               \\
Storage Practicality                                  & Good              & Best              & Worst              \\ \bottomrule
\end{tabular}
\label{tbl:summary}
\end{table}

The case studies demonstrate hydrogen's particular strengths in emissions reduction, with the light aircraft study showing significant decreases in nitrogen oxide and carbon monoxide emissions compared to conventional fuels. The military aircraft application revealed only modest efficiency losses—approximately 2.23\% in non-afterburning operation—while maintaining acceptable performance characteristics for high-demand operations.

It also becomes clear that hydrogen demonstrates superior environmental performance compared to conventional jet fuel, with significantly reduced emissions when produced through clean methods. The case studies show that even hydrogen produced from natural gas offers substantial emissions reductions compared to conventional aviation fuels. To that end, while hydrogen presents certain storage challenges, these prove more manageable than the integration issues posed by battery-electric systems. The military aircraft study particularly highlights how hydrogen's superior energy-to-mass ratio provides crucial advantages for high-performance applications where weight and space constraints are critical. 

Finally, the hybrid system tested in the light aircraft case study suggests a promising pathway for addressing reliability concerns, though with some impact on overall system efficiency. This approach maintains hydrogen's environmental benefits while providing operational redundancy. While our analysis focuses on technical feasibility, economic considerations will influence adoption timelines. Current hydrogen production costs (\$5-7/kg) and infrastructure requirements suggest initial implementation in high-value applications like military aviation, with broader commercial adoption following cost reductions through scale and technology improvements.

The primary implementation challenges center on storage system design and integration. Cryogenic storage requirements for liquid hydrogen necessitate specialized tank designs and careful consideration of their placement within aircraft structures. However, these challenges appear less severe than initially anticipated, particularly when compared to the fundamental limitations of battery-electric systems in aviation applications.

\section{Conclusion}\label{sec:conclusion}

The aviation industry's significant environmental impact necessitates the transition to sustainable fuel alternatives, with hydrogen emerging as a promising solution. This paper addressed the challenge of evaluating hydrogen's viability as an aviation fuel source by analyzing its efficiency, storage requirements, and practical implementations. Through comprehensive examination of case studies in light aircraft and military applications, we demonstrated that hydrogen-based systems can achieve performance metrics comparable to traditional fuels while significantly reducing emissions, particularly when compared to battery-electric alternatives. Our findings contribute to the field by establishing hydrogen's superiority over other green energy sources, supported by quantitative evidence from real-world implementations showing minimal efficiency losses and substantial emission reductions. However, limitations exist in the form of storage challenges that necessitate aircraft design modifications. Future research should address three specific areas: (1) development of lightweight cryogenic storage systems targeting a 15\% weight reduction over current technologies, (2) optimization of fuel cell systems for high-altitude operation above 30,000 feet, and (3) establishment of standardized fueling protocols for commercial aviation applications. These advances would facilitate hydrogen's widespread adoption while meeting stringent aviation safety requirements.

\bibliographystyle{IEEEtran}
\bibliography{refs}

\end{document}